# Network Coexistence Analysis of RIS-Assisted Wireless Communications

**Yajun Zhao[1,2], Xin Lv[1]**

[1]Beijing Institute of Technology, Beijing 100081, China;
[2]ZTE Corporation, Beijing 100029, China.

Corresponding author: Yajun Zhao (e-mail: zhao.yajun1@zte.com.cn).

**ABSTRACT** Reconfigurable intelligent surfaces (RISs) have attracted the attention of academia and industry circles because of their ability to control the electromagnetic characteristics of channel environments. However, it has been found that the introduction of an RIS may bring new and more serious network coexistence problems. It may even further deteriorate the network performance if these new network coexistence problems cannot be effectively solved. In this paper, an RIS network coexistence model is proposed and discussed in detail, and these problems are deeply analysed. Two novel RIS design mechanisms, including a novel multilayer RIS structure with an out-of-band filter and an RIS blocking mechanism, are further explored. Finally, numerical results and a discussion are given.

## I. INTRODUCTION

People have always dreamed of constantly controlling electromagnetic waves. The emergence of Maxwell's equations has enabled the ability of human beings to control electromagnetic waves to rapidly increase. However, due to the relatively fixed electromagnetic parameters of natural materials, the power to control electromagnetic waves is limited to transmitters and receivers. In recent years, reconfigurable intelligent surfaces (RISs) have attracted the attention of academia and industry since they can manipulate the electromagnetic characteristics of channel environments. Especially in the past one or two years, RISs have been developing rapidly in academic research and industrial scenarios. They are considered one of the key candidate technologies for 5G-Advanced [1]-[4] and 6G networks [5]-[7].

An RIS is a two-dimensional programmable meta-material composed of a large number of electromagnetic units that are arranged periodically. It can reconstruct the arrangement of the electromagnetic response structure of the meta-surface through the states of its switching elements, e.g., PIN diodes, varactor diodes, and liquid crystals, to realize changes in the electromagnetic response characteristics.

The electromagnetic unit structures can directly control electromagnetic waves in free space. A typical architecture of an RIS consists of three sub-layers and a controller [8]. These three sub-layers together constitute the functional layer of the RIS to support the functions of a dynamically tunable meta-surface. In this paper, the functional layer composed of these three sub-layers is regarded as an RIS layer. The geometric shape, size, direction, arrangement, reflection amplitude and phase shift of each electromagnetic unit can be reasonably designed to change the reconfigurable characteristics of the RIS elements, accordingly realizing the real-time reconstruction of the wireless environment. This digital programmable meta-surface is the basic functional layer of the traditional RIS referred to in this paper. More detailed concepts regarding the RIS layer are discussed in detail in section III.

The existing RIS research mainly focuses on the new challenges that classical communication problems have faced since the introduction of RISs, such as channel estimation [9]-[13] and beamforming [14]-[16], and these studies focus on channel models in single-network scenarios [17]-[20]. According to our limited search, except in one review article [21], we did not find any literature on RIS network coexistence scenarios. The coexistence of multiple networks is a traditional problem in practical wireless mobile communication networks. One of the main purposes of introducing an RIS in a wireless network is to use its ability to control the propagation of electromagnetic waves to overcome interference problems. However, we must face a sad reality; that is, the introduction of an RIS may bring about new and more serious network coexistence problems. If these new network coexistence problems cannot be effectively solved, the network performance will deteriorate further. The key challenge of multinetwork coexistence lies in the limited cooperation ability among networks because these networks are usually deployed and managed by different operators. In actual networks, the wireless signals incident on an RIS include both the "target signals" coming from the network to which the RIS belongs and the "nontarget signals" coming from neighbouring networks. Naturally, it is known that while optimizing the propagation characteristics of the target signals, this operation of the RIS will have unexpected effects on the nontarget signals.

**Our contributions**: Based on our previous studies [21], this paper makes two contributions: (1) to deeply analyse and model RIS network coexistence for the first time and (2) to further analyse and evaluate two novel RIS structures, including a novel multilayer RIS structure with an out-of-band filter and an RIS blocking mechanism.

The structure of the article is as follows. Section II discusses the proposed RIS network coexistence model in detail and deeply analyses the existing problems. In the third section, we further discuss two novel RIS design mechanisms, including a novel multilayer RIS structure with an out-of-band filter and an RIS blocking mechanism. The fourth part gives the numerical results and discussion. Finally, in the fifth part, a conclusion is drawn.

## II. SYSTEM MODEL AND PROBLEMS

### A. Traditional System Model

For the downlink of the RIS-assisted wireless communication system shown in FIGURE 1, we consider a node B (NB) with $M$ antennas and an RIS with $N$ elements serving a user with $K$ antennas. Let $H_{nb-ue} \in C^{K \times M}$ represent the direct channel between the user and the NB; $G_{nb-ris} \in C^{N \times M}$ is the channel between the RIS and the NB, and $H_{ris-ue} \in C^{K \times N}$ is the channel between the user and the RIS.

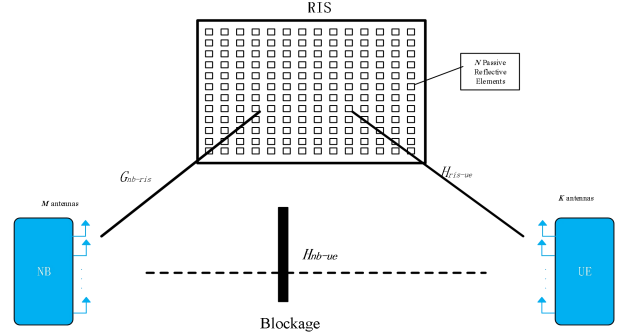

FIGURE 1a.

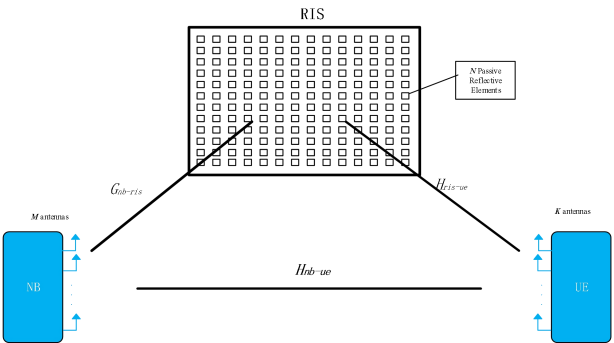

FIGURE 1b.

FIGURE 1. System model

The overall propagation channel $H_{total}$ between the NB and UE can be represented as follows (including two scenarios).

Scenario 1 concerns both the direct channel and the channel through the RIS (in FIGURE 1a):

$$H_{total} = H_{ris-ue} \Theta_{ris} G_{nb-ris} + H_{nb-ue} \qquad (1)$$

Scenario 2 involves only the channel through the RIS, while the direct channel is blocked (in FIGURE 1b):

$$H_{total} = H_{ris-ue} \Theta_{ris} G_{nb-ris} \qquad (2)$$

The received signal $Y$ at the NB can accordingly be represented as follows (also including two scenarios).

Scenario 1 concerns both the direct channel and the channel through the RIS (see FIGURE 1a):

$$Y = (H_{ris-ue} \Theta_{ris} G_{nb-ris} + H_{nb-ue})FX + W \qquad (3)$$

Scenario 2 considers only the channel through the RIS, while the direct channel is blocked (see FIGURE 1b):

$$Y = (H_{ris-ue} \Theta_{ris} G_{nb-ris})FX + W \qquad (4)$$

where $\Theta_{ris} = diag(\theta_1, \theta_2, \cdots, \theta_N)$ is the reflection matrix at the RIS with $\theta_i$ representing the reflection coefficient for the *n-th* RIS element, $F^{M \times 1}$ is the precoding matrix at the NB, and $W \in C^{U \times 1}$ is the noise at the UE. Note that $\theta_i$ can be further set as $\theta_i = \beta_i e^{j\varphi_i}$, where $\beta_i \in [0,1]$ and $\phi_i \in [0,2\pi]$ represent the amplitude and the phase of the *n-th* RIS element, respectively.

### B. RIS Network Coexistence Model and Problem Analysis

The traditional relay has complete RF units, which can employ bandpass filters to relay signals. Therefore, the relay only processes and relays the target signals and has little influence on the nontarget signals transmitted in adjacent frequency bands. In contrast, an existing RIS without RF units has no filtering function and usually has broadband tuning capability (i.e., a bandwidth of several GHz) [22][23]. The broadband tuning characteristic of an RIS is beneficial for wireless broadband communication and supports multiple bands [24]-[27], but it may lead to serious network coexistence problems [19]. From a traditional perspective, each element of an RIS can only be set with a single weighting coefficient and cannot be set with different weighting coefficients for different signals from different sub-bands within the frequency range tuned by the RIS [28]-[32]. The RIS with the tuning characteristic uses the same weighted coefficient matrix to tune all signals incident on it in a wider frequency band. Therefore, the existing RIS cannot perform optimal channel matching for more than one sub-band channel simultaneously. As shown in FIGURE 2, there are two overlapping networks, network A and network B (denoted by $N_A$ and $N_B$, respectively), which use adjacent frequency bands. The base station $nb_A$ belongs to network $N_A$, while the base station $nb_B$ belongs to network $N_B$. $ue_A$ and $ue_B$ are served by $nb_A$ and $nb_B$, respectively. The RIS ($RIS_A$) of network $N_A$ tunes the signal from network $N_A$ according to the channel. $RIS_A$ also tunes the signal from network $N_B$ within the same coverage area at the same time based on the same tuning coefficient matrix for the signal of network $N_A$. Then, the operation of $RIS_A$ causes a serious unexpected disturbance on the channel of the nontarget signal from network $N_B$, i.e., it leads to coexistence problems between networks $N_A$ and $N_B$. Notably, the tuning of $RIS_A$ to the signal of network $N_B$ is unexpected and may have serious impacts on the performance of network $N_B$.

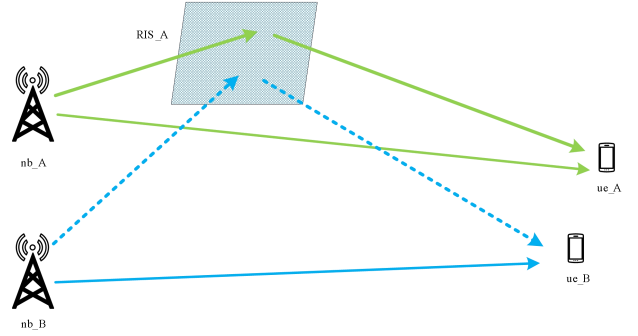

**FIGURE 2**. The coexistence of multiple RIS networks

#### 1) RIS NETWORK COEXISTENCE MODEL

Based on the above analysis, this section gives the coexistence model for an RIS network. Without loss of generality, assuming that there is a scattering path (the channel component of $nb-ris-ue$) and a direct path (the channel component of $nb-ue$), formula (3) can be modified as follows.

$$Y_{ue\_A} = (H_{ris\_A-ue\_A}\Theta_{ris\_A}G_{nb\_A-ris\_A} + H_{nb\_A-ue\_A})F_A X_{ue\_A} + W_{ue\_A} \quad (5)$$

$$Y_{ue\_B} = (H_{ris\_B-ue\_B}\Theta_{ris\_A}G_{nb\_B-ris\_A} + H_{nb\_B-ue\_B})F_B X_{ue\_B} + W_{ue\_B} \quad (6)$$

where $Y_{ue\_A}$ is the received signal at $ue_A$ served by base station $nb_A$ of network $N_A$, $Y_{ue\_B}$ is the received signal at $ue_B$ served by base station $nb_B$ of network $N_B$, and $\Theta_{ris\_A}$ is the optimized tuning coefficient matrix of $RIS_A$ for $ue_A$.

Because $RIS_A$ can only have one tuning state at a time, its coefficient matrix $\Theta_{ris\_A}$ that is suitable for the channel of $ue_A$ is also used to simultaneously tune the incident signals from network $N_B$. That is, as shown in formula (6), the tuning coefficient matrix for the signal of $ue_B$ is also $\Theta_{ris\_A}$.

#### 2) NONSTATIONARITY OF THE COEXISTING CHANNELS IN RIS NETWORKS

We can also analyse the channel of the nontarget signals from the perspective of statistical channel characteristics.

The artificial unexpected tuning of the RIS to the nontarget signal channel brings unexpected channel transients, thus losing the stationary characteristic of the original natural scattering channel. This leads to significant nonstationary characteristics in the channel of the nontarget signals.

The tuning coefficient matrix $\Theta_{ris\_A}$ is calculated based on the channel of $ue_A$, while the channels of $ue_A$ and $ue_B$ are independent and random. Therefore, an independent random relationship is also present between the channels of $\Theta_{ris\_A}$ and $ue_B$, which results in unexpected random tuning changes in the channel of $ue_B$. For example, the channel $H_{ue\_B}(t_0)$ of $ue_B$ is measured at time $t_0$, and $ue_B$ sends its signal at time $t_1$ when the tuning coefficient matrix of $ris_A$ is $\Theta_{ris\_A}$. Then, channel $H_{ue\_B}(t_1)$, which is unexpectedly tuned by $\Theta_{ris\_A}$, may be quite different from $H_{ue\_B}(t_0)$. These issues lead to channel mismatches and may lead to serious performance degradation, especially when the subpath reflected by the RIS accounts for a high proportion of the total signal energy.

Furthermore, this setting can be divided into the following two cases.

(1) Case 1: High-load scenarios

Generally, in a case with a high load, adjacent consecutive time slots are dynamically allocated to different UEs. Accordingly, the tuning matrix of an RIS is dynamically changed to serve the different UEs, which means that the unexpected tuning of nontarget signals is very dynamic. When the time interval $T_{ris}$ of the change of the RIS tuning matrix coefficient is less than the time $T_{meas}$ of the CSI measurement ($T_{ris} < T_{meas}$), it is difficult to accurately and reliably obtain the CSI.

(2) Case 2: Low-load scenarios

In a case with a low load, the carrier of the target network is idle most of the time. During the idle period of a network, it is only necessary to change the tuning coefficient matrix semistatically to meet the imposed coverage requirements, that is, broadcasting synchronization signals and system information, receiving uplink random access, etc. Since the condition $T_{ris} > T_{meas}$ can be met most of the time in this case, the CSI of nontarget signals can be accurately obtained. The channel transient only occurs in the time slot where the target signals are scheduled to be transmitted. The impact on the average throughput of nontarget signals is small because the channel is unexpectedly tuned only in sparsely distributed time slots.

With a limited literature search, we found that the spatial-temporal nonstationary characteristics of the nontarget signal channel are very different from those of traditional nonstationary channel models. The existing studies on channel nonstationary characteristics assume the natural scattering characteristic [33][34]. In contrast, in the RIS coexistence scenario, the nonstationarity of the nontarget signal channel is mainly caused by the unexpected abnormal tuning of the RIS. At least two factor characteristics lead to channel nonstationarity: (1) unexpected independent processes and (2) abnormal meta-surface scattering characteristics, which are different from those of natural materials. The nontarget signal channel matrix described in this paper is tuned by an independent random matrix, which cannot be modeled by the traditional nonstationary channel model, nor can it be estimated by the channel estimation methods for the traditional nonstationary channel. This paper provides a mechanism to suppress unexpected tuning, reduce the influence of an unexpected tuning matrix and reduce the variance of the channel nonstationary fluctuations to solve this kind of nonstationary channel problem.

3) CHANNEL CAPACITY INFLUENCE CAUSED BY THE COEXISTENCE OF RIS NETWORKS

In this section, we briefly analyse the influence of the unexpected tuning of nontarget signals by an RIS from the perspective of channel capacity.

For a random channel, assuming that both the transmitter and receiver have perfect CSI, the formula of MIMO channel capacity $C$ constructed using the singular value decomposition (SVD) transmission structure is as follows.

$$C \approx \sum_{i=1}^{n_{\min}} E\left[\log(1+\frac{SNR}{n_{\min}}\lambda_i^2)\right] \quad (7)$$

When $n_r = 1$,

$$C = E\left[\log(1+SNR\sum_{i=1}^{n_t}|h_i|^2)\right] \quad (8)$$

where $n_{\min}$ is the minimum number of antennas in the sending nodes or the receiving nodes, i.e., the number of data streams sent by the MIMO subspace in parallel; $\lambda_i$ is the power coefficient of the $i$ th data stream; $SNR$ is the signal-to-noise ratio of the receiver under the assumption of power normalization; and $h_i$ is the channel between the $i$ th antenna and the receiving antenna.

As described above, it is difficult for the transmitter to accurately obtain the CSI of a signal unexpectedly tuned by the RIS. Especially when the channel component tuned by the RIS contributes a high proportion of the propagation channel of nontarget signals, unexpected RIS tuning makes the obtained CSI completely unreliable. In this case, the transmitter must assume unknown CSI. When the CSI at the transmitter is unknown, a typical transmitter precoding matrix is the identity matrix. In this case, the MIMO channel capacity $C$ is

$$C = E\left[ log(1 + \frac{n_r}{n_t} SNR \sum_{i=1}^{n_t} |h_i|^2 ) \right] \qquad (9)$$

When $n_r = 1$,

$$C = E\left[ log(1 + SNR / n_t \sum_{i=1}^{n_t} |h_i|^2 ) \right] \qquad (10)$$

where $n_t$ is the number of antennas in the transmitting node; $n_r$ is the number of antennas in the receiving node; $h_i$ is the channel between two antennas of the transmission and the receiver ; and $SNR$ is the signal-to-noise ratio of the receiver.

Compared with perfectly known CSI at the transmitter, unknown CSI at the transmitter incurs the loss of the beamforming gain. For example, a transmitter with 2 transmission antennas loses 3 dB.

4) EXAMPLE - A SPECIAL SCENARIO

For some special areas, only one main scatter can provide electromagnetic wave scattering to achieve coverage. As shown in FIGURE 3a, the signal of base station A ($nb_A$) belonging to network $N_A$ cannot cover a target area through the natural scattering of the wall itself. Base station B ($nb_B$), belonging to network $N_B$, can cover the target area exactly. To overcome the problem of covering the hole in this area, network $N_A$ deploys an RIS on the wall to tune its signal propagation. The coverage problem of network $N_A$ can be effectively overcome, but the coverage of network $N_B$ is destroyed due to the unexpected tuning of the RIS for its signal propagation. As shown in FIGURE 3b, the beam from network $N_B$ can no longer cover the target area, which causes this area to become the coverage hole of network $N_B$. In addition, if the CSI of the signal from network $N_B$, i.e., the so-called "nontarget signals", cannot be accurately obtained, it is difficult for the nontarget signals to guarantee the quality of service.

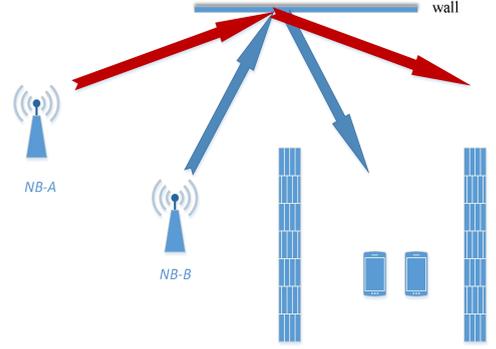

FIGURE 3a.

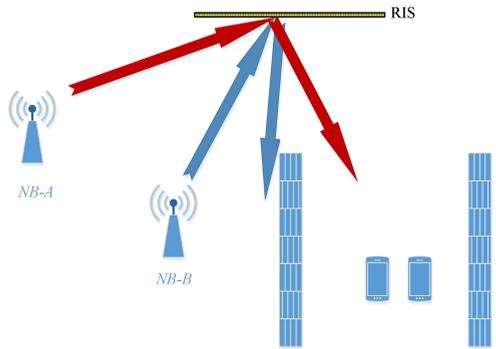

FIGURE 3b.

FIGURE 3. Example: A special case for overcoming blind coverage

III. INNOVATION MECHANISMS FOR THE RIS NETWORK COEXISTENCE PROBLEM

As mentioned earlier, the coexistence problem of an RIS network is caused by the fact that the RIS uses the same coefficient matrix to tune the target signals and nontarget signals, which are simultaneously incident on its surface. We need to find mechanisms to reduce or eliminate the unexpected tuning of nontarget signals. From the perspective of eliminating this influence, a mechanism can be designed so that nontarget signals will not be incident on the RIS surface and will not be tuned unexpectedly. From the perspective of reducing the influence, it is possible to maximally reduce the proportion of unexpected nontarget signal energy tuning. In this section, two novel RIS design mechanisms proposed by us [21], including a novel multilayer RIS structure with an out-of-band filter and an RIS blocking mechanism, are further explored. Among them, the former adopts the idea of eliminating influence, while the latter adopts the idea of reducing influence. The advantages and possible negative effects of these two mechanisms are comprehensively analysed and evaluated.

*A. A Novel Multilayer RIS Structure with an Out-of-Band Filter Layer*

1) STRUCTURES AND MODELS

Various groups have reported on multilayer meta-surfaces and multilayer dielectric substrates for higher-order bandpass resonances [35]-[39]. However, in our previous research, a novel multilayer RIS structure with an out-of-band filter was proposed for the first time and used to solve the coexistence problems of RIS networks [21]. On the basis of our previous works, this paper conducts a more in-depth theoretical analysis and performance evaluation of the multilayer RIS structure with an out-of-band filter to solve the problems of RIS network coexistence.

Arming at the problem of unexpected tuning in different frequency coexistence scenarios, a type of RIS with a multilayer meta-surface structure was proposed in our previous article [21]. Without loss of generality, let us take an RIS with a double-layer meta-surface structure as an example. The first layer of the RIS uses a bandpass filter with a fixed coefficient meta-surface. The meta-surface constituting the first layer is a transmission-type meta-surface. It only allows signals in the target band to pass through, while the signals in the adjacent nontarget band (out-of-band signals) are filtered. The second layer of the RIS is a conventional programmable meta-surface structure that can realize typical programmable RIS functions. The programmable meta-surface of the second layer only tunes the target signal, since the nontarget signal has been filtered by the first layer. It should be noted that the first layer of the reflective RIS with a double-layer structure filters the out-of-band signals twice (see FIGURE 4). In another works, the filter layer of the RIS performs filtering once when the signal is incident and once when it is reflected. In this case, the final filtering coefficient is the square of the unidirectional filtering coefficient. For a refraction RIS with a double-layer structure, the incident signal is filtered only once (refer to FIGURE 5a). Therefore, the preferred structure of the refraction RIS involves the design of out-of-band filter layers on the upper and lower surfaces of the RIS structure, i.e., an RIS with a triple-layer structure, which can achieve the same filtering effect as that of the reflection-type RIS (refer to FIGURE 5b). Notably, we can design an RIS with more filter layers to achieve a better filtering effect. However, more filter layers also cause challenges, such as complexity, cost, and volume issues, as well as even more negative impacts on the target signal. It is necessary to strike a balance between these challenges and the filtering performance (note that unless otherwise stated, the rest of this article only discusses the reflection RIS).

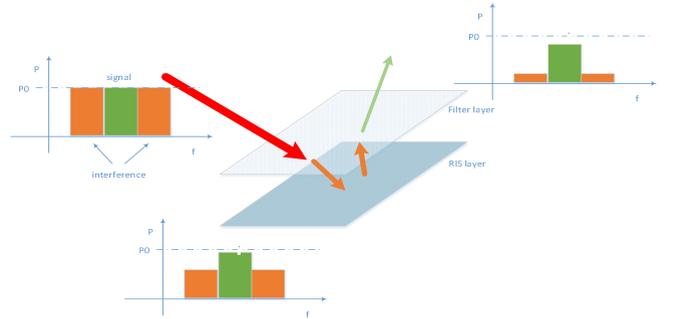

**FIGURE 4**. Out-of-band filtering using an RIS with a double-layer structure (reflecting the RIS)

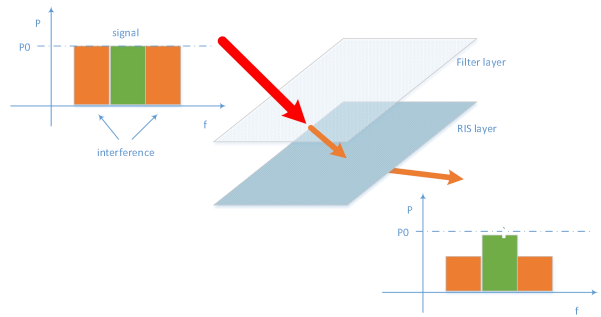

**FIGURE 5a.**

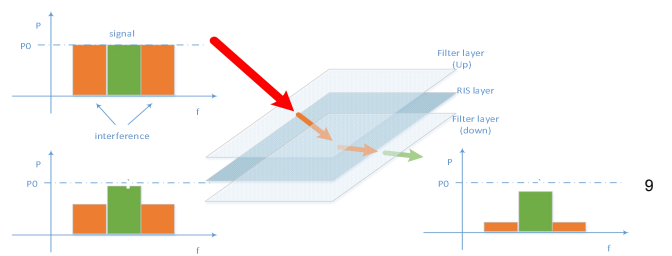

FIGURE 5. Out-of-band filtering using an RIS with a triple-layer structure (refracting the RIS)

It should be noted that when filtering nontarget signals, the target signals may also be influenced. For an RIS with a bandpass filter structure, the performance index of the out-of-band filter is generally inconsistent with the guarantee of the in-band index. On the premise of ensuring the tuning performance of the in-band target signal, improving the performance index of the out-of-band filter requires increased complexity and a higher cost. However, the tuning of a general RIS is quantified with finite bits while taking complexity and cost into account (for example, 1~3 bits). Considering that the quantization accuracy of the RIS itself is relatively low, in principle, the out-of-band performance index may be reduced to some extent.

From the perspective of the implementation mechanism, the filter layer of the RIS can be designed to absorb or scatter the nontarget signals. If it is designed to perfectly absorb the nontarget signals, no unexpected tuning will occur, but it will lead to energy loss in the nontarget signal. Since the equivalent size of the antenna aperture of an RIS is usually larger than that of a traditional filter, the energy loss caused by signal absorption is not ignored. In special cases, the RIS scattering path may be the only signal propagation path (refer to FIGURE 2), or the main scatterer surfaces of the propagation environment may be widely deployed by the RIS. If the energy of the nontarget signals is perfectly absorbed, communication failure occurs. Therefore, for the above situation, the optimal design goal of the bandpass filter is to maximize the transmission of the target signals and the scattering of the nontarget signals.

Without loss of generality, suppose that there are two neighbouring base stations $NB_A$ and $NB_B$, which belong to networks $N_A$ and $N_B$, respectively; the $RIS_A$ and $RIS_B$ deployed in neighbouring positions belong to these two base stations; and $NB_A$ and $NB_B$ use the adjacent frequency bands to transmit signals (in FIGURE 3). The signals from $NB_A$ are incident on $RIS_A$ and $RIS_B$. The signals from $NB_A$ are optimally tuned by $RIS_A$ but unexpectedly tuned by $RIS_B$. If $RIS_B$ has an out-of-band filter layer, the unexpected tuning of the $NB_A$ signals in adjacent frequency bands can be effectively suppressed. Correspondingly, the signals from $NB_B$ undergo similar processing. Then, the expression of the $ue_A$ signal served by $NB_A$ and the $ue_B$ signal served by $NB_B$ can be modified from formula (3), as shown below.

$$Y_{ue\_A} = (H_{ris\_A-ue\_A}\Theta_{ris\_A}G_{nb\_A-ris\_A} + H_{nb\_A-ue\_A})F_A X_{ue\_A} + (H_{ris\_B-ue\_A}\beta\Theta_{ris\_B}G_{nb\_A-ris\_B})F_A X_{ue\_A} + W_{ue\_A}$$
(11)

$$Y_{ue\_B} = (H_{ris\_B-ue\_B}\Theta_{ris\_B}G_{nb\_B-ris\_B} + H_{nb\_B-ue\_B})F_B X_{ue\_B} + (H_{ris\_A-ue\_B}\beta\Theta_{ris\_A}G_{nb\_B-ris\_A})F_B X_{ue\_B} + W_{ue\_B}$$
(12)

In formula (11), $Y_{ue\_A}$ refers to the received signal of $ue_A$ served by the base station $nb_A$, $\beta \in (0,1)$ refers to the filter coefficient of $ris_A$, $\Theta_{ris\_A}$ refers to the optimal tuning matrix of $ris_A$ for the signal of $ue_A$, and $\Theta_{ris\_B}$ refers to the optimal tuning matrix of $ris_B$ for the signal of $ue_B$.

For convenience of description, in formulas (11) and (12), the "filter coefficient $\beta$" is defined as the proportion of energy remaining after a signal is filtered. It should be noted that the filtering coefficient of the formula is squared, i.e., $(\sqrt{\beta})^2 = \beta$, because the signal undergoes out-of-band filtering twice when entering and exiting the filtering meta-surface structure of the RIS. We can see that $\Theta_{ris\_B}$ unexpectedly tunes the signal of $ue_A$, but the unexpected tuning effect is suppressed by the weighting coefficient $\beta$. Accordingly, formula (12) represents the signal of $ue_B$ served by $NB_B$. It is necessary to design a reasonable filter coefficient $\beta$ to effectively improve the coexistence performance of adjacent frequency bands.

2) ANALYSIS OF CHANNEL CHARACTERISTICS

It is assumed that the channel model is a multipath channel model with multiple scatters, and RISs with adjacent-frequency multilayer filter structures are deployed on the surfaces of the main scatterers. While optimizing the transmission path of a target signal, the RIS filters the nontarget signals with the adjacent frequency that are incident on the surface of the RIS. Without loss of generality, assume that the subpath component $H_k$ of the nontarget signal is incident on a scatter surface $S_k$. If a traditional RIS is deployed, the channel matrix $H_{non-target}$ of the filtered nontarget signal can be expressed as

$$H_{non-target} = GH_k + \sum_{i=1, i \neq k}^{N} H_i \quad (13)$$

where $G$ is the beamforming gain of the RIS and $N$ is the number of subpaths in the nontarget signal channel.

It is assumed that the proposed multilayer RIS with an out-of-band filter layer is arranged on the surface of scatterer $S_k$. The filters can be further divided into two types: absorption-mode filters and scattering-mode filters. The channel characteristics of RISs with these two types of filter layer structures will be analysed (note: without loss of generality, for the convenience of the analysis, we assume that the scattering response of the signal component naturally scattered by the RIS filter layer surface is consistent with that of the original scattering surface).

(1) Case A: RIS with an absorption-mode filter

If an RIS with an absorption-mode filter is deployed and the absorption coefficient of the nontarget signal is $\beta$, the channel matrix $H_{non-target\_f}$ of the filtered nontarget signal can be expressed as follows.

$$H_{non-target\_f} = \beta H_{sub-k1} \Theta_{target} H_{sub-k2} + \sum_{i=1, i \neq k}^{N} H_i \quad (14)$$

where $0 \leq \beta \leq 1$, $H_{sub-k1}$ is the channel between the scatter and UE, $H_{sub-k2}$ is the channel between the scatter and NB, $\Theta_{target}$ is the corresponding RIS tuning matrix for the target signal, $H_i$ is the subchannel of the NBs directly connected to the UE, and $N$ is the number of subchannels of the NBs directly connected to the UE.

Formula (14) shows that $H_{non-target\_f}$ loses the energy of part of the subpath $\sqrt{1-\beta} H_k$, and only the $\sqrt{\beta} H_k$ part is tuned by the RIS.

Few-layer meta-surfaces have demonstrated novel functionalities such as perfect absorption [40]-[42]. If the nontarget signal is perfectly absorbed, i.e., $\beta = 0$, the channel matrix $H'_{non-target\_f}$ of the filtered nontarget signal can be rewritten as follows.

$$H'_{non-target\_f} = \sum_{i=1, i \neq k}^{N} H_i \quad (15)$$

Furthermore, if the nontarget signal only passes through the RIS path, the components of other subpaths are very weak and can be ignored; i.e., $H_i \approx 0$. In this case, the channel of the nontarget signal is expressed as

$$H_{non-target\_f} \approx 0 \quad (16)$$

Then, the channel condition of the nontarget signal is very poor, and the network is no longer able to provide communication services for the UE.

(2) Case B: RIS with a scattering-mode filter

In this case, an RIS with a scattering-mode filter is deployed, and the scattering coefficient is $\Theta_0$. Without loss of generality, it is assumed that the scattering characteristics of the RIS for nontarget signals are the same as those of the original natural scatters, except for the energy coefficient ($\beta \leq 1$). $\beta \leq 1$ means that a certain percentage of the energy transmission of the nontarget signal falls on the RIS surface and is unexpectedly tuned; then, the channel $H_{non-target\_f}$ of the nontarget signal can be expressed as

$$H_{non-target\_f} = (\sqrt{\beta})^2 H_{re\_sub1} \Theta_0 H_{re\_sub2} + (\sqrt{1-\beta})^2 H_{re\_sub1} \Theta_{target} H_{re\_sub2} + \sum_{i=1, i \neq k}^{N} H_i \quad (17)$$

where $\Theta_0$ is the natural scattering response of the scatter, $H_{re\_sub1}$ is the channel between the scatter and the UE, $H_{re\_sub2}$ is the channel between the scatter and the NB, $\Theta_{target}$ is the corresponding RIS tuning matrix for the target signal, $H_i$ is the subchannel of the NBs directly connected to the UE, and $N$ is the number of subchannels of the NB directly connected to the UE.

In a special case, the nontarget signals are perfectly scattered, i.e., $\beta = 1$, and the channel of the nontarget signals is exactly the same as that without RIS deployment.

$$H_{non-target\_f} = H_{non-target} \qquad (18)$$

## B. A Novel RIS Blocking Mechanism

### 1) STRUCTURES AND MODELS

Our previous study put forward a novel RIS blocking mechanism [43][21]. In [43], an RIS with a large size was divided into several subblocks to serve different UEs separately. In [21], we first proposed an RIS blocking mechanism to solve the RIS network coexistence problem. Based on our previous research, this paper conducts a more in-depth theoretical analysis and performance evaluation of the RIS blocking mechanism to solve the RIS network coexistence problem.

The basic idea of the RIS blocking mechanism is that the incident signals of different UEs can be assigned to different subblocks by using an independent coefficient matrix; these blocks are used for the simultaneous tuning and beamforming of these signals. From the perspective of the UE source, the mechanism is used for multi-UE scheduling when the UEs come from the same network. The mechanism is used for multinetwork coexistence when the UEs come from different networks. For the RIS network coexistence scenario described in this paper, it needs to be assumed that the RIS can only be blocked in a static or semistatic manner since it is difficult to dynamically coordinate between base stations, especially base stations coming from different operators. Under the assumption that the RIS can only be divided in a static or semistatic way, it is necessary to deploy an appropriate RIS antenna scale and design a reasonable blocking ratio to ensure the RIS performance as much as possible while satisfying the coexistence performance. In addition, when analysing the performance of the RIS block, the article [21] only considers the gains of subblocks allocated to the target UE and does not consider the influence of other blocks on the unexpected tuning of UE signals.

As shown in FIGURE 6, a signal is incident on several subblocks instead of one subblock in an RIS because the size of a beam spread is usually larger than that of the RIS subblock or even larger than the whole RIS. Therefore, the different signal components are usually tuned by multiple subblocks simultaneously. It can be seen that the signal components falling on their own RIS subblock will be optimized and tuned as needed, while the components falling on other subblocks will be tuned unexpectedly. Without loss of generality, assuming that an RIS is divided into two subblocks and that the $ue_A$ signal is set as the target signal, formula (3) can be modified as follows.

$$Y_{ue\_A} = \sqrt{\beta}(H_{ris\_sub1-ue\_A}\Theta_{ris\_A}G_{nb\_A-ris\_sub1} + H_{nb\_A-ue\_A})F_A X_{ue\_A} +$$

$$\sqrt{(1-\beta)}(H_{ris\_sub2-ue\_A}\Theta_{ris\_B}G_{nb\_A-ris\_sub2})F_A X_{ue\_A} + W_{ue\_A}$$
(19)

where $Y_{ue\_A}$ is the received signal of $ue_A$ served by base station $nb_A$, $\beta$ is the energy proportion of the $ue_A$ signal incident on its own RIS subblock, $(1-\beta)$ is the energy proportion of the $ue_A$ signal incident on another RIS subblock, $\Theta_{ris\_A}$ is the optimal tuning matrix of the RIS subblock for the $ue_A$ signal, and $\Theta_{ris\_B}$ is the optimal tuning matrix of the RIS subblock for the $ue_B$ signal.

According to formula (19), the energy of the $ue_A$ signal is unexpectedly tuned by $\Theta_{ris\_B}$ with a ratio of $(1-\beta)$. Therefore, naturally, we can learn that the main factors of the RIS blocking mechanism should at least design an appropriate RIS size and an appropriate block scale factor $\beta$.

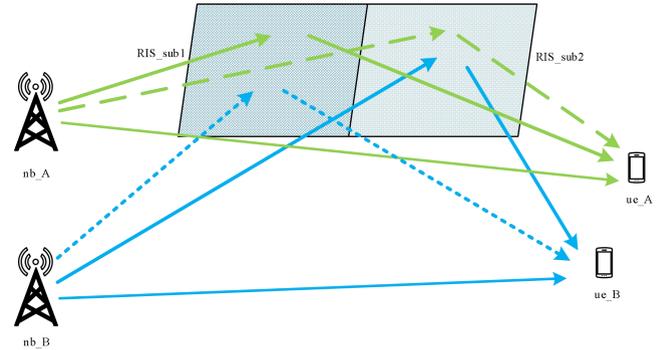

**FIGURE 6.** RIS blocking supports network coexistence

### 2) CHANNEL CHARACTERISTIC ANALYSIS

When the RIS is divided into two subblocks, the tuning coefficient matrix $\Theta_{ris}$ consists of two groups of elements, one for the tuning coefficient $\theta_{a\_ij}$ of $ue_A$ and one for the tuning coefficient $\theta_{b\_kl}$ of $ue_B$.

$$\Theta_{ris} = \begin{bmatrix} \theta_{a\_11} & \theta_{a\_21} & \cdots & \theta_{b\_i1} \\ \theta_{b\_21} & \cdots & \cdots & \cdots \\ \cdots & \cdots & \cdots & \cdots \\ \theta_{a\_1i} & \cdots & \cdots & \cdots \end{bmatrix} \quad (20)$$

Without loss of generality, we simply divide the RIS into an upper subblock $A$ and a lower subblock $B$ as an example. Then, the tuning coefficient matrix $\Theta_{ris}$ can be expressed as $\Theta'_{ris}$.

$$\Theta'_{ris} = \begin{bmatrix} \Theta_{ue\_A} & \\ & \Theta_{ue\_B} \end{bmatrix} \quad (21)$$

where $\Theta_{ue\_A}$ is the optimal tuning coefficient matrix of the RIS subblock allocated to $ue_A$ and $\Theta_{ue\_B}$ is the optimal tuning coefficient matrix of the RIS subblock allocated to $ue_B$. $\Theta'_{ris}$ tunes all signals incident on it. Therefore, $\Theta_{ue\_B}$ unexpectedly tunes the signal of $ue_A$ incident on it, and $\Theta_{ue\_A}$ unexpectedly tunes the signal of $ue_B$ incident on it.

The above analysis is mainly aimed at scenarios that satisfy the traditional typical far-field channel conditions. As discussed in reference [16], a certain proportion of scenarios can meet the near-field channel conditions because typical RISs are large in size and widely deployed in networks. For scenarios satisfying the near-field channel conditions, the transmitter is closer to the RIS. If the beam is narrow and the main lobe expansion of the signal is less, most of the beam energy will fall on one of the RIS subblocks. In this way, the narrow beams of different UEs can be adapted to fall on different subblocks of the RIS, and the tuning coefficient matrix of each RIS subblock can be adapted for different UEs. Only a small proportion of the signal's energy may fall on other subblocks and be tuned unexpectedly. In this case, a higher performance upper bound can be obtained by using the RIS blocking mechanism.

In brief, although the use of the RIS blocking mechanism may reduce the "good" network performance to a certain extent, it can greatly improve the "bad" network performance (refer to FIGURE 10 for numerical results). In near-field scenarios with narrow beams, the blocking mechanism can yield better performance. It is necessary to carefully design an appropriate RIS size and an appropriate block scale factor $\beta$ and pursue overall performance enhancements for multiple coexisting networks.

## IV. NUMERICAL RESULTS AND DISCUSSION

In the above chapters, we have conducted in-depth theoretical analyses of the coexistence problems of RIS networks and two corresponding solutions. Based on these analyses, this section carries out numerical simulations for some typical cases using Monte-Carlo simulation.

In each simulation, we assume that there are two networks, one NB for each network and one UE for each NB. The channel is assumed to be a far-field model. Unless specified otherwise, we assume that the RIS employs a UPA with $M = M_x \times M_y$; we fix $M_x$ = 20 and increase $M_y$ = [4,8,16,32]*2; the NB adopts a UPA with $N = N_x \times N_y$, $N_x$ =8 and $N_y$ = 4; and the UE adopts $K$ = 1. The carrier frequency is set to 28 GHz. The normalized power $p$ is 1, and the variance of the Gaussian white noise is $\sigma$ = 3.16e-11.

### A. Evaluation of the Impact of the Unexpected Tuning of an RIS on the Performance of Neighbouring Networks

First, we use the beam pattern to qualitatively analyse the performance influence. Without loss of generality, FIGURE 7 shows the beam pattern in free space, in which the target user is located on the normal path of the beam.

The channel of the nontarget signal is independent of those of the target signal. The angle between the position of the nontarget signal and the normal beam of the target signal is random, and the nontarget users are randomly distributed with an angle range of $(-\frac{\pi}{2}, \frac{\pi}{2})$. As shown in FIGURE 7, the signal quality will be very good if the nontarget UE happens to be located near the peak. On the other hand, if the nontarget UE happens to be in the trough, the signal will be very poor, even causing communication interruptions.

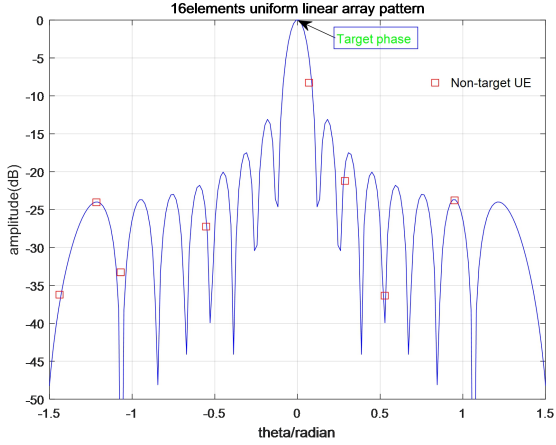

FIGURE 7. 16-elements uniform linear array pattern

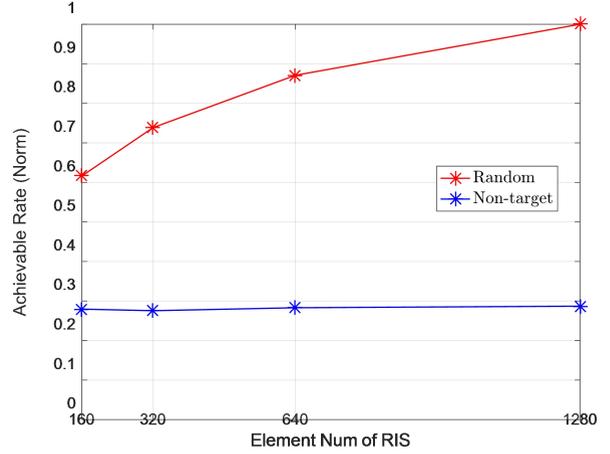

FIGURE 8. Performance impacts caused by unexpected RIS tuning

("Random": random phase scattering of incident electron waves by natural scatterers; "Nontarget": unexpected RIS tuning of the incident electromagnetic wave)

Without loss of generality, scenarios with only one RIS scattering path are simulated here. Suppose that two independent networks (networks $N_A$ and $N_B$) are present and that network $N_A$ is deployed with the RIS. The signals of network $N_B$ may be unexpectedly tuned by the RIS of network $N_A$. Assume that the signals incident on the RIS will be scattered in a random phase before the RIS is deployed on the surface of the scatterer. FIGURE 8 shows the performance comparison regarding the nontarget signals in the case with random phase scattering and unexpected RIS tuning. FIGURE 8 depicts the achievable data rate versus the number $M$ of RIS elements, where the transmission power $p$ is fixed and $M$ varies. The "Random" curve indicates the performance achieved when the incident electromagnetic wave is randomly scattered; the "Nontarget" curve indicates the performance achieved when the incident electromagnetic wave is unexpectedly tuned by the RIS. It can be seen from the curves in FIGURE 8 that unexpected RIS tuning leads to a significant decrease in the achievable data rate of the nontarget UE. With an increase in the number of antenna arrays, the performance does not improve.

### B. Evaluation of the Performance of the Multilayer RIS Structure with Out-of-Band Filtering

Using an RIS with a filter structure, nontarget signals incident on the RIS surface can be filtered out to effectively suppress unexpected tuning. FIGURE 9 depicts the achievable data rate versus the number $M$ of RIS elements, where the transmission power $p$ is set to 50 dBm and $M$ varies. Without loss of generality, it is assumed that the RIS filter layer scatters the nontarget signals incident on it with a random phase (please refer to formula (17)), and the nontarget signal are tuned by the RIS in all time slots. FIGURE 9 shows the performance achieved with different filter coefficients. When the nontarget signals are perfectly scattered, that is, the filter coefficient factor $\beta = 1$ (please refer to formula (18)), the best performance is attained. When the scattering coefficient is 0.5, the influence of unexpected tuning on the nontarget signals can also be reduced. The simulation results show that the multilayer filtering mechanism can effectively solve the RIS network coexistence problems caused by the unexpected tuning of nontarget signals.

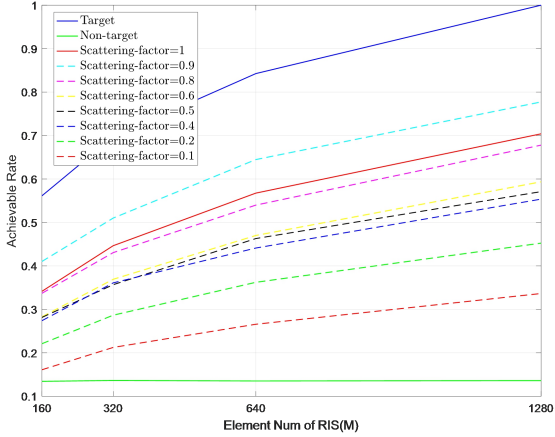

FIGURE 9. Nontarget signal performance of the RIS with a multiple-layer structure

("Target": the RIS tunes the signals using perfect CSI; "Nontarget": the RIS unexpectedly tunes the incident electromagnetic wave using the normal RIS; "scattering factor=1": perfect scattering; "scattering factor=0.9/.../ 0.1": the energy scattering coefficient scattered by the multilayer RIS)

FIGURE 9 shows the simulation results of the special case in which the RIS performs tuning in all time slots, that is, the simulation results obtained under the full buffer service model. Generally, networks have different service loads, which are reflected in the probabilities of RIS tuning in different time slots. Without loss of generality, we assume that the scattering coefficient is $\beta = 0.8$ and simulate the process under different RIS tuning probabilities. We assume that the RIS performs random phase scattering on the signal when it does not perform tuning. FIGURE 10 shows the performance of the proposed RIS with a multilayer structure under different loads, that is, under different tuning probabilities. The results show that the performance impact caused by the unexpected tuning of the traditional RIS increases with increasing network load. The proposed RIS with a multilayer structure has good performance under different network loads.

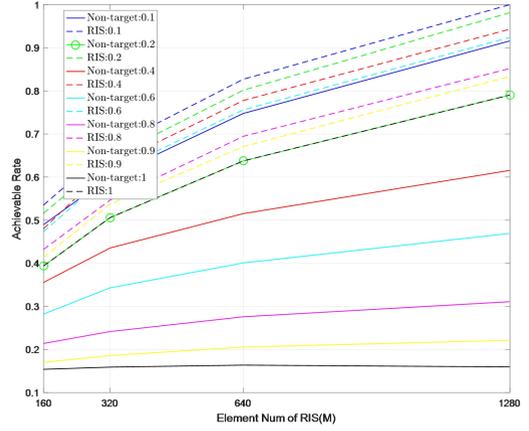

FIGURE 10. Performance of the RIS with a multilayer structure under different loads

("Nontarget": unexpected RIS tuning of the incident electromagnetic wave using the normal RIS; "RIS": the multilayer RIS; "x =0.1/0.2/.../0.9": RIS tuning probability)

To better evaluate the performance of the proposed multilayer RIS with a filter layer, we evaluate the performance it achieves when deployed in different locations. We assume that other conditions are constant and that the fixed scattering coefficient is $\beta = 0.8$; only the distance between the NB and RIS panel changes. FIGURE 11 shows that the RIS is deployed at different distances from the NB, and the RIS can provide good gains.

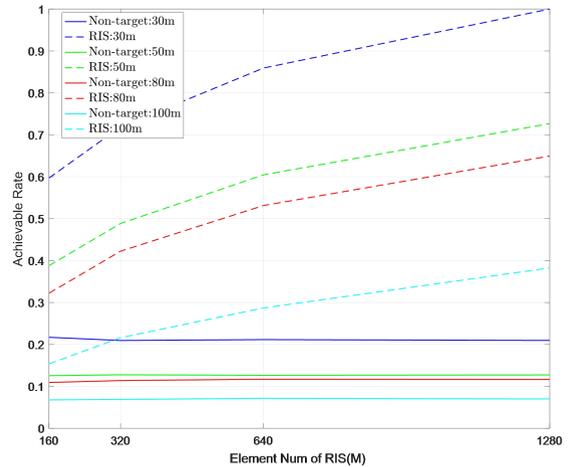

FIGURE 11. Performance of the RIS with a multilayer structure when deployed at different distances from the NB

("Nontarget": unexpected RIS tuning of the incident electromagnetic wave using the normal RIS; "RIS": the multilayer RIS; "x =30 m/50/80 m/100 m": the RIS is deployed at different distances from the NB)

FIGURE 12 further provides the performance achieved when two RISs are deployed in a cell. Without loss of generality, we also assume that the scattering coefficient is 0.8. The results of FIGURE 12 show that when multiple RISs are deployed in a cell, the proposed multilayer RIS attains better performance.

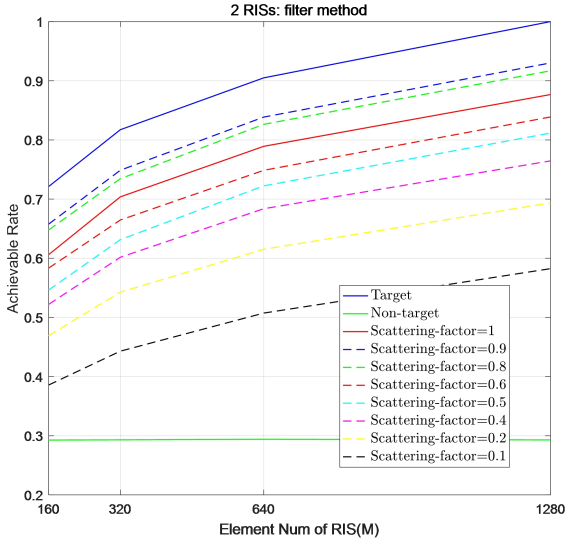

FIGURE 12. Performance of the RIS with a multiple-layer structure when two RISs are deployed in a cell

("Target": the RIS tunes the signals using perfect CSI; "Nontarget": the RIS unexpectedly tunes the incident electromagnetic wave using the normal RIS; "scattering factor=1": perfect scattering; "scattering factor=0.9/.../ 0.1": the energy scattering coefficient scattered by the multilayer RIS)

## C. Simulation and Analysis of the RIS blocking mechanism

Without loss of generality, it is assumed that two networks share one RIS, and the RIS blocking mechanism is simulated and evaluated. According to formula (19) for the RIS blocking mechanism, the tuning coefficient of one RIS subblock matches the target signal, while the tuning coefficient of the other subblock matches the traditional nontarget signal. In the simulation, we simulate different coefficients $\beta$, that is, different signal energy ratios incident on the two subblocks. Without loss of generality, we can assume that the energy ratio of the signals incident on the two subblocks is equal to the proportion of the block sizes, where the coefficient of unexpectedly tuned signal energy is $\beta$, and we let the proportion $\beta \in \{0.1, 0.2, 0.4, 0.5, 0.6, 0.8, 0.9\}$. FIGURES 10-11 depict the achievable data rate versus the number $M$ of RIS elements, where the transmission power $p$ is fixed and $M$ varies. It is assumed that the nontarget signals are tuned by the RIS in all time slots. From the curve in FIGURE 10, we can see that the proposed mechanism can greatly improve the performance attained for nontarget signals, but it also causes the target signal performance to be reduced to some extent. Because the RIS blocking mechanism reduces the effective antenna aperture of the target signal, the RIS subblocks assigned to other users will unexpectedly tune the target signal. However, in FIGURE 11, the sum data rate of the target UE and nontarget UE, i.e., the target signal and nontarget signal shows that this RIS blocking mechanism can achieve a considerable sum rate and improve the overall performance of multiple networks.

In addition, the RIS with a blocking mechanism can use the tuning ability of the RIS to cover some corner areas that cannot be covered by natural scattering. In other words, this RIS blocking mechanism can guarantee basic coverage performance in RIS network coexistence scenarios and can overcome some extreme situations, such as coverage holes.

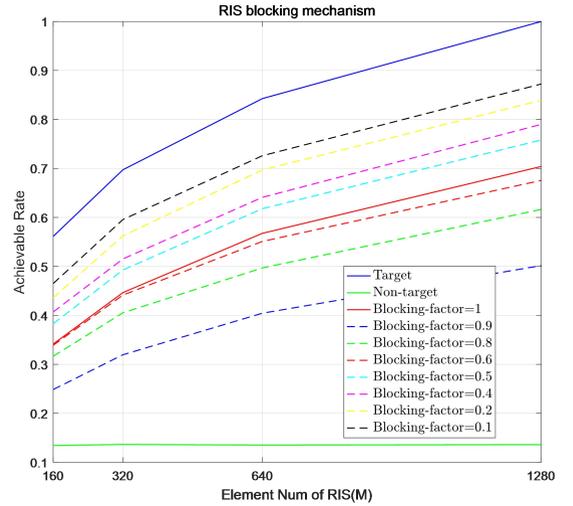

FIGURE 12. Performance of the RIS blocking mechanism

("Target": RIS signal tuning using perfect CSI; "Nontarget": unexpected RIS tuning of the incident electromagnetic wave using the normal RIS; "Blocking factor={0.9..., 0.1}": the size ratio $\beta$ of an RIS for the nontarget signal)

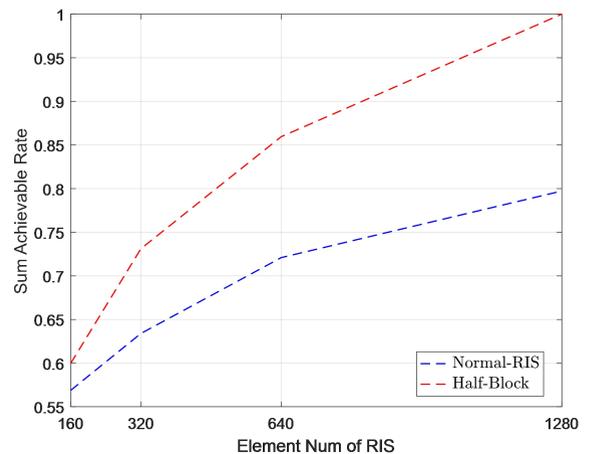

FIGURE 13. Sum achievable rate of the target UE and nontarget UE

("Normal RIS": sum data rate of the target and nontarget signals using the normal RIS; "Half-block": sum data rate of the target and nontarget signals using the RIS with a blocking mechanism)

Similar to the above evaluation of the multilayer RIS, more scenarios are also examined for the block RIS structure.

The results in FIGURES 10 and 11 show the simulation results of the RIS in the special case in which all time slots are tuned, that is, the simulation results obtained under the full buffer service model. Generally, networks have different service loads, which are reflected in the probabilities of RIS tuning in different time slots. Without loss of generality, we assume that the RIS is equally divided into two sub-blocks (i.e., $\beta = 0.5$), which are used for the target signal of the serving cell and the traditional nontarget signal of the neighbouring cell, respectively. FIGURE 14 shows the performance of the proposed RIS blocking mechanism under different loads, that is, different tuning probabilities.

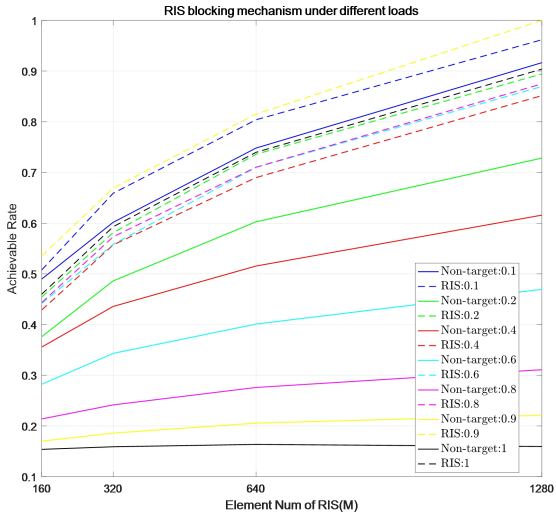

FIGURE 14. Performance of the RIS blocking mechanism under different loads

("Nontarget": unexpected RIS tuning of the incident electromagnetic wave using the normal RIS; "RIS": the multilayer RIS; "x =0.1/0.2/.../0.9": RIS tuning probability)

To better evaluate the performance of the proposed RIS blocking mechanism, we evaluate the performance attained by the RIS deployed in different locations. We assume that the other conditions remain unchanged; the RIS is fixedly and equally divided into two subblocks (i.e., $\beta = 0.5$), which are used for the target signal of the serving cell and the traditional nontarget signal of the neighbouring cell, and only the distance between the NB and RIS panel changes.

FIGURE 15 shows that the RIS is deployed at different distances from the NB, and the RIS can provide good gains.

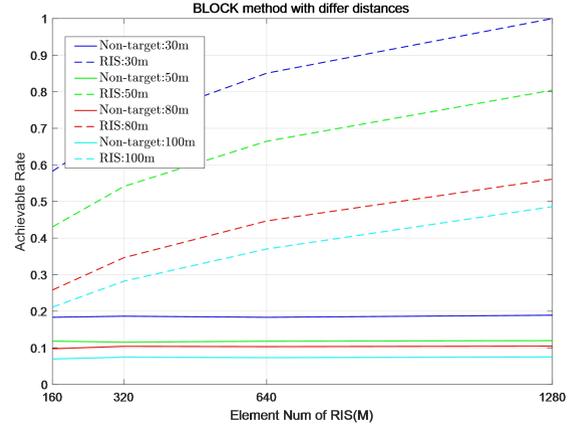

FIGURE 15. Performance of the RIS using blocking mechanism deployed at different distances from the NB

("Nontarget": unexpected RIS tuning of the incident electromagnetic wave using the normal RIS; "RIS": the multilayer RIS; "x =30 m/50/80 m/100 m": the RIS is deployed at different distances from the NB)

FIGURE 16 further provides the performance achieved when two of the proposed RISs are deployed in a cell. Without loss of generality, we also assume that each RIS is fixedly and equally divided into two sub blocks (i.e., $\beta = 0.5$), which are used for the target signal of the serving cell and the traditional nontarget signal of the neighbouring cell. The results in FIGURE 16 show that when multiple RISs are deployed in a cell, the RISs yield better performance.

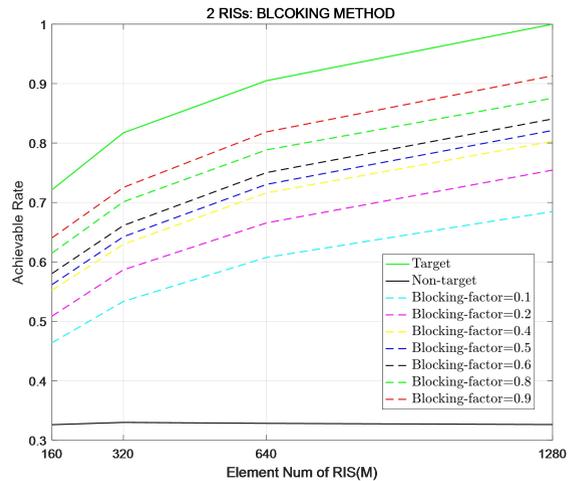

FIGURE 16. Performance achieved when two RISs using blocking mechanism are deployed in a cell

("Target": RIS signal tuning using perfect CSI; "Nontarget": unexpected RIS tuning of the incident electromagnetic wave using the normal RIS; "Blocking factor={0.9..., 0.1}": the size ratio $\beta$ of an RIS for the nontarget signal)

## V. CONCLUSION

This paper discusses the coexistence of wireless networks after introducing an RIS and provides the channel model obtained under the coexistence scenario. The analysis and evaluation show that the deployment of an RIS has serious negative impacts on network performance if the coexistence problem cannot be effectively solved. We further carry out a theoretical analysis and a numerical simulation evaluation on the two newly proposed RIS structures, proving that they can effectively solve the problem of RIS network coexistence.

From the above analysis, it can be seen that the two mechanisms provided in this paper are only applicable to limited scenarios. Additional typical scenarios require further analysis and evaluation. For more complex situations, it is necessary to further optimize the parameters of the two proposed mechanisms or introduce other possible mechanisms to achieve a balance between performance and cost. In addition, from the point of view of spectrum reuse, network coexistence can include two modes, network coexistence on the same frequency band (also known as cochannel coexistence) and network coexistence in different frequency bands (also known as adjacent channel coexistence). The solution proposed in this paper is mainly applicable to coexistence scenarios with different frequency bands, and a more appropriate solution for coexistence scenarios involving the same frequency band remains to be further studied.

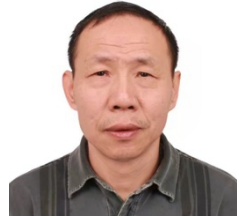

**Xin Lv** received the B.S., M.S., and Ph.D. degrees in electronic engineering from the Beijing Institute of Technology (BIT), in 1982, 1988, and 1993, respectively. Since 1982, he has been with BIT, as a Lecturer, an Associate Professor, and a Professor, and now he is retired.

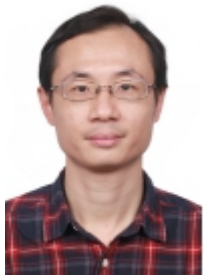

**Yajun Zhao** has B.E. and master's degrees. Since 2010, he has acted as a radio expert in the wireless advanced research department at the ZTE Corporation. Prior to this, he worked for Huawei on wireless technology research in the wireless research department. Currently, he is mainly engaged in research on 5G standardization technology and future mobile communication technology (6G). His research interests include reconfigurable intelligent surface (RIS), spectrum sharing, flexible duplex, CoMP, and interference mitigation.